# Electronic origin of the orthorhombic *Cmca* structure in compressed elements and binary alloys

Valentina F. Degtyareva

Institute of Solid State Physics RAS, Chernogolovka, Moscow district, 142432, Russia;
E-Mail: degtyar@issp.ac.ru

**Abstract:** Formation of the complex structure with 16 atoms in the orthorhombic cell, space group *Cmca* (Pearson symbol *oC*16) was experimentally found under high pressure in the alkali elements (K, Rb, Cs) and polyvalent elements of groups IV (Si, Ge) and V (Bi). Intermetallic phases with this structure form under pressure in binary Bi-based alloys (Bi-Sn, Bi-In, Bi-Pb). Stability of the *Cmca* - *oC*16 structure is analyzed within the nearly free-electron model in the frame of Fermi sphere – Brillouin zone interaction. A Brillouin-Jones zone formed by a group of strong diffraction reflections close to the Fermi sphere is the reason for reduction of crystal energy and stabilization of the structure. This zone corresponds well to the 4 valence electrons in Si and Ge and leads to assume a *spd*-hybridization for Bi. To explain the stabilization of this structure within the same model in alkali metals, that are monovalent at ambient conditions, a possibility of an overlap of the core and valence band electrons at strong compression is considered. The assumption of the increase in the number of valence electrons helps to understand sequences of complex structures in compressed alkali elements and unusual changes in their physical properties such as electrical resistance and superconductivity.

## 1. Introduction

Experimental studies of elements under high pressure reveal a great variety of structural transformations summarized in review papers (see [1-3] and literature cited therein). At ambient pressure the most representative structures for elements are close-packed structures: body-centred cubic (bcc), face-centred cubic (fcc), and hexagonal close-packed (hcp), typical for metals. Elements on the right-hand side of the Periodic table – main group IV - VII semi-metals and non-metals – crystallize in open-packed structures following the 8-N coordination rule. These elements tend to transform under pressure to a metallic state with closely packed structures. On the left-hand side of the Periodic table the group 1 and II elements (the alkali and alkali-earth elements) are good metals at ambient pressure that crystallize in the closely packed structures. Under pressure, however, they show an anomalous behaviour transforming to open-packed structures.

Two different trends from the both sides of the Periodic table are illustrated by the data in Table 1 that shows the structural sequences for Si, Ge and Bi (upper panel) and for the heavy alkali elements (lower panel). These elements are chosen for our considerations because under pressure they all go in their pathways through the same structure – orthorhombic with 16 atoms in the cell, space group *Cmca*, Pearson symbol *oC*16. This type of structure is unknown for any element or compound at ambient pressure and was defined first for Cs and Si under pressure [4,5] and then for other elements included in Table 1. Theoretical calculations of the band structure energy for *Cmca* phases of Si and Cs have demonstrated that "structural similarities in Cs-V and Si-VI are not reflected in similarities of the electronic structures" [6]. For Si, a nearly free-electron behavior has been shown, while for Cs-V, dominate *d*-orbital components due to $s-d$ electron transfer have been found [6].

In this paper formation and stability of the *oC*16-*Cmca* structure is discussed within the Brillouin zone – Fermi sphere interaction with application of the Hume-Rothery mechanism [7-9]. A complex structure, like γ-brass $Cu_5Zn_8$-*cI*52, reduces its crystal energy by formation of the Brillouin zone planes close to the Fermi sphere satisfying the electron concentration rule. Following this model it is necessary to assume for compressed alkali metal not only $s-d$ transfer but also further overlap of valence band and upper core electrons [10]. This assumption was made to account for an unusual behavior under pressure of the alkali element sodium [11]. Na transforms above 1 Mbar to the low-symmetry, open structure *oP*8, which is similar to the phase AuGa with 2 electrons per atom. It was suggested for Na-*oP*8 an electronic transfer from core into the valence band. Similar suggestion is necessary to assume for alkali elements in *oC*16-*Cmca* phase as is discussed in this paper.

**Table 1**. Sequences of structural transformations on pressure increase
for elements with the *oC*16 structure

| | |
|---|---|
| **Si** | 12    13    16    38    42    80<br>*cF*8 → β-Sn, *tI*4 → *oI*4 → *hP*1 → ***oC*16** → hcp → fcc < 250 GPa<br>Diamond    white tin |
| **Ge** | 11    75    85    102    160<br>*cF*8 → β-Sn, *tI*4 → *oI*4 → *hP*1 → ***oC*16** → hcp < 180 GPa<br>Diamond    white tin |
| **Bi** | 2.5    2.7    7.7<br>*hR*2 → *mC*4 → h-g → bcc < 220 GPa<br>→ ***oC*16** (>483K) → |
| **K** | 11.6    20    54    90    96<br>bcc → fcc → h-g → *oP*8 → *tI*4 → ***oC*16** < 112 GPa<br>25    35<br>→ *hP*4 → |
| **Rb** | 7    13    17    20    48<br>bcc → fcc → *oC*52 → h-g → *tI*4 → ***oC*16** < 70 GPa |
| **Cs** | 2.4    4.2    4.3    12    72<br>bcc → fcc → *oC*84 → *tI*4 → ***oC*16** → dhcp < 223 GPa |

Notes: The numbers above the arrows show the transition pressures in GPa, reported in experimental measurements on pressure increase. The crystal structures of the phases are denoted with their Pearson symbols, apart from the most common metallic structures bcc, fcc, and hcp. "h–g" stands for a host-guest structure. For experimental data see [1-3] and literature cited therein.

## 2. Results and Discussion

The crystal structure $oC16$-$Cmca$ belongs to the family of structures described by Pearson [12] as planar square-triangle nets of atoms, assigned as $3^2434$ where "3" specifies a triangle, "4" a square, and the superscript shows a multiplicity. These nets are alternated with square $4^4$ nets. Layers $3^2434$ may be stacked either antisymmetric or with a shift on 1/2 of a basic axis. Examples of this family represent the compounds $CuAl_2$ ($tI12$), $CoGe_2$ ($oC24$), $PdSn_2$ ($tI48$), $PdSn_3$ ($oC32$) and others [12]. Interesting example is an $oC20$ phase found for $PdSn_4$, $PtSn_4$ and $AuSn_4$ [13]. This type of structure with the space group $Aba2$ has some similarities with the discussed $oC16$-$Cmca$ structure having nearly same axial ratios but slightly different stacking layers: four $3^2434$ nets alternate by two $4^4$ nets, whereas in $oC16$-$Cmca$ the numbers of nets are two and four, respectively, as considered below.

*2.1. Structural characteristics of oC16-Cmca phases*

The conventional unit cell contains 16 atoms in two crystallographically non-equivalent positions 8f $(0,y,z)$ and 8d $(x,0,0)$ in Wyckoff notation of the $Cmca$ space group. A feature of the structure $oC16$-$Cmca$ is $3^2434$ nets formed by atoms in the 8f position with squares lying over the cell corners and cell base centers or over the midpoints of the cell edges and alternated by two square nets $4^4$ (atoms in the 8d position) shifted with respect to each other on a half of the *b* axis as shown in Figure 1. The configuration of atoms in the $3^2434$ nets implies a rigid atomic arrangement with atomic positions $y$ and $z$ (8f) close to 0.167 and 0.333, respectively, providing $y + z \approx 0.5$.

**Figure 1.** Crystal structure $oC16$-$Cmca$ of Si-VI. **(a)** A perspective view. The atomic positions 8f and 8d are denoted as red and yellow, respectively. **(b)** A projection along the *a*-axis for nets at x=0.5 (red) and x = 0.218 and 0.282 (yellow). The structure can be viewed as an alternating sequence of planar layers formed by square-triangle nets and puckered nearly square layers. The shortest interatomic distances are marked by dark lines (modified from [4]).

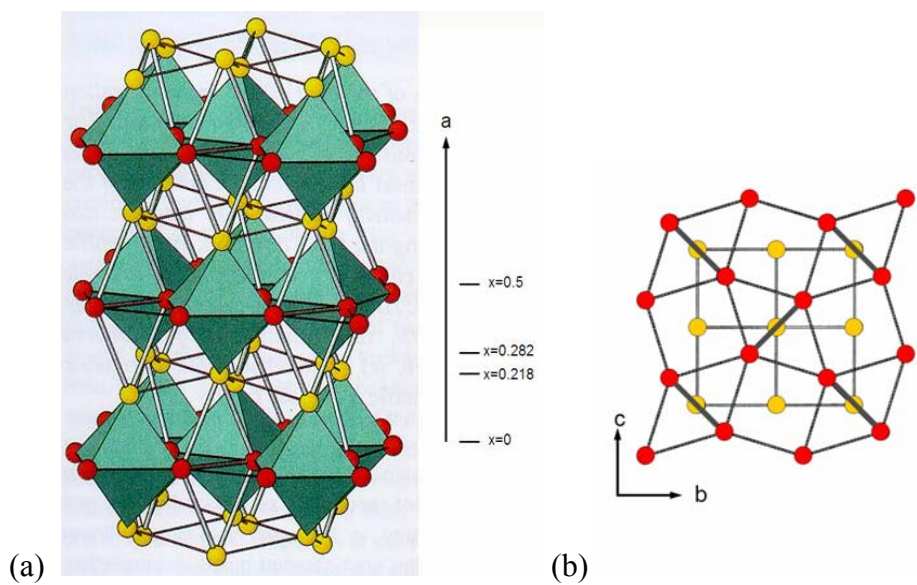

The data of all *oC*16-*Cmca* phases found at present are summarized in Table 2. All the structures have very similar axial ratios ($b \approx c \approx 1$, $a/b \approx a/c \approx \sqrt{3}$) and very similar atomic positions. This indicates that the structure is defined by certain criteria related to the electronic structure.

**Table 2**. Structural details of all the known *Cmca-oC*16 phases, formed under pressure in the pure elements Cs, Rb, Si, Ge and Bi, and in Bi-alloys with In, Sn and Pb.

| Phase | Si-VI 38.4GPa [5] | Ge 135GPa [14] | Cs-V 12GPa [4] | Rb-VI 48.1GPa [15] | K-VI 112GPa [16] | Bi-IV 3.2GPa 465K [17] | $Bi_{40}Sn_{60}$ 5.8GPa [18] | $Bi_{80}Pb_{20}$ quenched from 1.7 GPa [19] | $Bi_{70}In_{30}$ quenched from 2.0 GPa [19] |
|---|---|---|---|---|---|---|---|---|---|
| $a$ (Å) | 8.024 | 7.885 | 11.205 | 9.372 | 8.032 | 11.191 | 10.7345 | 11.209 | 11.236 |
| $b$ (Å) | 4.796 | 4.655 | 6.626 | 5.5501 | 4.753 | 6.622 | 6.3270 | 6.672 | 6.555 |
| $c$ (Å) | 4.776 | 4.651 | 6.595 | 5.5278 | 4.716 | 6.608 | 6.3115 | 6.658 | 6.531 |
| $V_{at}$(Å$^3$) | 11.49 | 10.67 | 30.60 | 17.97 | 11.25 | 30.61 | 26.79 | 31.12 | 30.07 |
| $V/V_o$ | 0.565 | 0.470 | **0.264** | **0.194** | **0.150** | 0.864 | 0.882 | 0.905 | 0.921 |
| $a/c$ | 1.680 | 1.695 | 1.699 | 1.689 | 1.703 | 1.693 | 1.701 | 1.684 | 1.721 |
| $b/c$ | 1.004 | 1.001 | 1.005 | 1.004 | 1.004 | 1.002 | 1.002 | 1.002 | 1.004 |
| $d_{short}$(Å) | 2.340 | 2.315 | **3.235** | **2.764** | **2.317** | 3.275 | 2.985 | 3.242 | 3.146 |
| y (8f) | 0.173 | 0.173 | 0.164 | 0.170 | 0.173 | 0.170 | 0.167 | 0.172 | 0.170 |
| z (8f) | 0.327 | 0.328 | 0.313 | 0.318 | 0.327 | 0.320 | 0.333 | 0.328 | 0.330 |
| x (8d) | 0.216 | 0.218 | 0.218 | 0.211 | 0.216 | 0.210 | 0.212 | 0.214 | 0.212 |

Some data for *Cmca-oC*16 phases in the Table 2 are essentially different, namely volume compression, $V/V_o$, that is especially significant for alkali metals (indicated in bold). The shortest interatomic distances for alkali metals Cs, Rb and K (Table 2) correspond to atomic radii 1.62, 1.38 and 1.16 Å, respectively. It is interesting to compare these values to the ionic radii given by Kittel [20], 1.67, 1.48 and 1.38 Å, or with ionic radii for coordination number 10 given by Shannon [21], 1.81, 1.66 and 1.59 Å, respectively. At this degree of compression it is necessary to consider ionic core overleap and a transfer of core electrons into the valence band, as was theoretically proposed earlier [10].

*2.2. Electronic origin of stability for oC16 phases: Hume-Rothery effects*

Examination of stability of the *oC*16 structure is based on the concept of the Hume-Rothery mechanism and consists of construction of the Brillouin planes close to the Fermi sphere defined within the nearly free-electron model [22]. We consider first the *oC*16 phase in *sp*-elements and alloys and then in alkali elements.

2.2.1. Group IV elements Si and Ge with the *oC*16 structure.

Elements Si and Ge are *sp*-elements of group IV and therefore have 4 valence electrons per atom. Figure 2 (a) shows diffraction pattern for Si-*oC*16 with indication of position $2k_F$ for z = 4 that locates just above a group of strong diffraction peaks selected for construction of the Brillouin-Jones zone (left) [22]. The inscribed Fermi sphere corresponds to $2k_F$ for z = 4 and is in contact with 36 planes; zone filling by electron states $V_{FS}/V_{BZ}$ equals ~93% as for the classical Hume-Rothery phase γ-brass $Cu_5Zn_8$. For the spherical Fermi surface calculated distances to the Bragg planes (131), (113), (421) and (511) are slightly below the value of $k_F$ with ratios 1.032, 1.028, 1.014 and 1.004, respectively. In real case there is a degree of deformation of the Fermi sphere by the contact with the Brillouin boundary, so-called "truncation" factor. This was initially suggested by Sato and Toth for the electronic nature of long-period superlattices in alloys like Cu-Au and similar systems [23] where the value of this factor was found to be within 5%.

In the middle of Figure 2 (a) a Brillouin zone is shown that is constructed by the planes of the first group of the strong reflections lying within the Fermi sphere. It is remarkable, that due to particular ratios of cell axes this polyhedron is very symmetrical with hexagonal cross-sections ⊥ *c** and in planes *a***c** and *b***c**. The symmetry of the BZ polyhedron formed by first strong Bragg planes is reflected on the electrostatic energy (see by Harrison [24]).

2.2.2. Structure *oC*16 in Bi and Bi-alloys.

Crystal structure of high-pressure, high-temperature phase of bismuth at ~4GPa and ~500 K was recently solved by single-crystal synchrotron X-ray diffraction [17] with characteristics given in Table 2. Experimental observations of this Bi phase have a long story since 1958 [25]. Pressure-temperature ranges for this Bi phase, assigned now as Bi-IV, are from ~2.2 GPa to ~5.5 GPa above 450 K. Room temperature phase Bi-III, stable at pressures 2.7 – 7.7 GPa has a complex incommensurate host-guest structure [26]. Phase boundary Bi-III – Bi-IV is nearly horizontal (~ 450 K) and both phases transform to Bi-V with body-centered cubic structure at higher pressure. There is a small volume difference between Bi phases III and IV and different structure may be expected from the difference in valence electron bands. Formally, the group V element Bi has outer $s^2p^3$ electron configuration. Under pressure-temperature conditions one may expect that hybridization of *s*, *p* and *d* levels occurs due to lowering of the higher empty *d*-band, as discussed by Pearson (see Ref. [12], p. 230-232).

For stability of the *oC*16 structure 4 valence electrons are necessary and this count relates to the assumption of *spd* hybridization. It should be noted that for considerations of the Fermi sphere radius within the model of FS-BZ interactions only *sp* electrons are taken into account as structure controlled electrons. The same assumption on *spd* hybridization for Bi in Bi-based alloys with the

*oC*16 structure is necessary to obtain the electron per atom count equal to 4 which stabilizes this structure.

Intermediate phases in Bi-Pb and Bi-In alloys were obtained under pressure-temperature conditions and quenched to ambient pressure at liquid nitrogen temperature [27]. Diffraction patterns of both phases $Bi_{80}Pb_{20}$ and $Bi_{70}In_{30}$ were similar, however rather complex to be solved 40 years ago. Development of new experimental methods of X-ray diffraction under pressure and new programs of structure solutions allowed defining *oC*16-*Cmca* structure in Si and Cs and other elements as indicated in Table 2. Based on these findings a description of structure was found for similar phases in $Bi_{80}Pb_{20}$ and $Bi_{70}In_{30}$ alloys [19] and was also suggested for the Bi-IV phase. Experimental confirmation of this suggestion was obtained recently [17].

In the Bi-Sn alloys an intermediate high-pressure phase was observed *in situ* under pressure [18]. It is remarkable that the X-ray diffraction data of $Bi_{40}Sn_{60}$ phase reveals the *oC*16-*Cmca* structure with the evidences of site-ordering. Additional diffraction peaks (for example, the difference reflection 111) indicate preferred occupation of the constituent atoms Sn and Bi in positions 8f and 8d, respectively. To extend this observation one can expect formation of ordered *oC*16-*Cmca* structures in some binary systems under pressure on condition that electron concentration rule will be satisfied.

**Figure 2.** Simulated diffraction patterns (left) and constructed Brillouin-Jones zones (right) **(a)** for Si-*oC*16 and **(b)** for Cs-*oC*16 with the data from Table 2. Indices hkl for reflections selected for BZ constructions are indicated on the patterns. The position of $2k_F$ is shown by dotted (red) line. See discussion in the text.

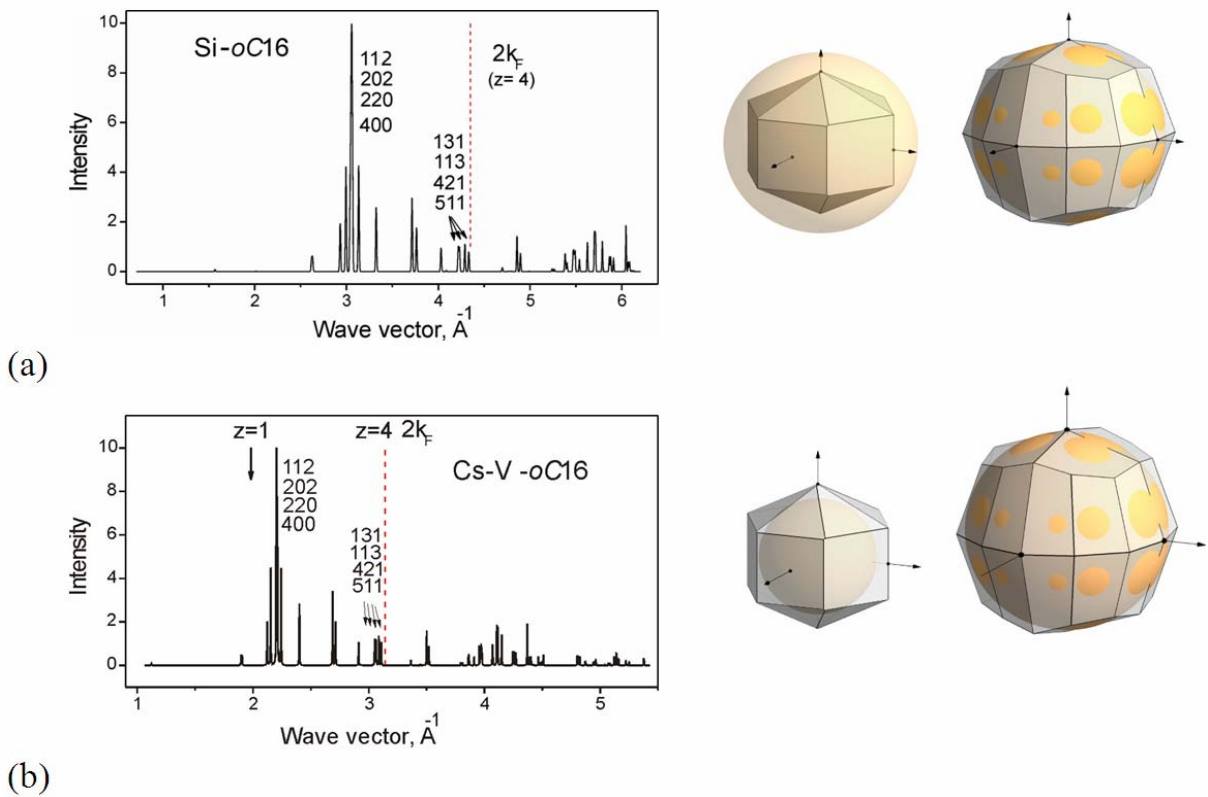

(a)

(b)

2.2.3. Structure *oC*16 in alkali elements: core ionization.

Phases with *oC*16-*Cmca* structure appear in Si and Ge under pressure within the sequence *cF*8 → *tI*4 → *hP*1 → *oC*16 → *hcp* that follows an increase in packing density and coordination number 4 → 6 → 6+2 → 10,11 → 12. This pathway is in general tendency for the elements on the right-hand side of the Periodic table to adopt the more close-packed structures. Opposite trend was found for the elements on the left-hand side – a transformation under compression to open-packed structures. Alkali elements with the bcc structure at ambient conditions which transforms to the fcc on pressure increase (CN = 8 and 12, respectively) undergo a transformation to low-coordinated structures. For K, Rb and Cs the structure before *oC*16 is *tI*4 with CN = (4+4), so that there is a turnover at transition *tI*4 → *oC*16 where CN starts to increase again and for Cs it attains CN = 12 at the transition to the double hexagonal close-packed structure, *dhcp*.

Unexpected behavior of alkali elements on compression tolerates radical changes in valence electron bonding and an electronic level overlap and hybridization of upper core electrons and valence band should be considered. This assumption is supported by the large compression and short interatomic distances in *oC*16 structure as is analyzed above in section *2.1*. On Figure 2(b) diffraction pattern Cs-*oC*16 is shown (left) with the calculated $2k_F$ positions shown for two possible numbers of valence electrons, z = 1 and 4. The groups of strong reflections near these positions are used for BZ construction (middle and right). The relation of the FS and BZ for the case z = 1 does not satisfy the Hume-Rothery conditions that supports the suggestion in increase of the valence electron counts. The Brillouin zone in Figure 2(b) (right) for z = 4 is similar to that of Si-*oC*16 in Figure 2(a).

*2.3. Structure stability and correlation with properties within the Hume-Rothery mechanism*

Stability of Hume-Rothery phases in copper-based alloys and other binary systems of noble-metals and *sp* polyvalent metals have been explained by the mechanism of Brillouin zone – Fermi-sphere interactions [7,8]. Hume-Rothery effects are found to become apparent for wide variety of material /physical objects as for structurally complex alloy phases, quasicrystals and their approximants [28,29]. Liquid and amorphous metals may be ascertain to obey the Hume-Rothery rules [30,31].

Fermi-sphere – Brillouin zone interactions should increase under compression providing minimization of electron band energy which depends on volume as $V^{-2/3}$ whereas the electrostatic energy depends as $V^{-1/3}$ (see discussion in [2]). The *oC*16-*Cmca* structure considered in this paper satisfies the criteria of Hume-Rothery mechanism for the elements Si and Ge with 4 valence electrons and there is no satisfaction of the Hume-Rothery mechanism for alkali metals if considered monovalent. If the *oC*16 structure in the alkali metals is considered to be stabilized by the Hume-Rothery mechanism, it is essential to assume the increase in the valence electron count to z = 4 in these metals due to the valence band overlap with the outer core electrons.

From another point of view atomic interactions can be analyzed using the electron localization function (ELF), which describes chemical bonding in real space [32,33]. The result of this approach is that a large amount of valence charge in the open-packed structures resides within the interstitial regions corresponding to the formation of 'electrides' in which the interstitial electrons form the anions. Thus an elemental phase under pressure is considered as "pseudobinary ionic compound" [34,35]. These papers report the observation of the hexagonal phases of NiAs-*hP*4 type in Na and K (above 180 GPa and 35 GPa, respectively), the type that is widely occurring in binary phases (see by Pearson [12]). The formation of this phase formed in a binary system of *sp*-metals can be well explained using the Hume-Rothery mechanism with an appropriate number of valence electrons. There is a great variety of complex phases based on *hP*4 formed by distortions, vacancies and superlattices to satisfy Hume-Rothery effects [36]. These considerations can help in understanding of structural variety observed in alkali elements under different pressure-temperature conditions.

Structural properties of materials in many instances correlate with their physical properties. Fermi-sphere – Brillouin zone interactions and degree of filling of the Brillouin zone by electron states define considerable the properties of material like electrical resistivity, optical reflectivity, superconductivity and others. With high degree of BZ filling the free Fermi surface area becomes small which is accompanied by an increase in density of state below the $k_F$. The consequences of this FS-BZ configuration are the decrease in reflectivity, increase in resistivity, appearance and increase in superconductivity, as discussed by Hirsch [37, 38].

For the phases with the *oC*16-*Cmca* structure in Cs and Si, experiments show a relatively high electrical resistance and an increase in $T_c$ of superconducting state (see discussion in [2]). Electrical resistance measurements under pressure available for Cs found a sharp increase at the transition *tI*4 – *oC*16 near ~13 GPa. For Cs superconductivity emerges in *tI*4 at ~11 GPa and riches $T_c$ ~1.3 – 1.5 K in *oC*16 at pressure 12-14 GPa [39]. For Si superconducting state was observed at the transition to the metallic state with β-Sn (*tI*4) and in the subsequent *oI*4 and *hP*1 phases with $T_c$ ~ 7 - 8 K at pressures 12 – 20 GPa, with a decrease of $T_c$ with pressure within the existence of the simple hexagonal (*hP*1) phase. It is noticeable that after reaching the minimum of 3.3 K at 37 GPa $T_c$ starts to increase again and reaches 4.9 K at 40 GPa in the *oC*16 phase followed by a drop in the *hcp* phase [40]. The observed relatively high $T_c$ values in the *oC*16 phases can be attributed to the higher degree of filling of the Brillouin-Jones zone by the electronic states.

The drop of resistance for Bi at the transition of the ambient pressure semimetallic phase to Bi-II - *mC*4 at 2.5 GPa is well known from its magnitude, sharpness, and reversibility and make it an ideal pressure calibration point [25]. Further transition to Bi-III (host-guest) at 2.7 GPa leads to an increase of resistance and the high temperature phase Bi-IV - *oC*16 has resistance that is a little lower than for Bi-III. At higher pressures the transition IV-V is accompanied by a moderately sharp drop in resistance. This behavior of resistance in Bi can be related now to established crystal structures of high pressure phases. Relatively high resistance of Bi-IV is in agreement with our considerations of FS-BZ model for *oC*16 structure. In addition, the relatively high $T_c$ values were observed for quenched *oC*16 phases in $Bi_{80}Pb_{20}$ and $Bi_{70}In_{30}$ alloys, 8.7 K and 7.9 K, respectively [27].

In the Bi-Sn alloys of the composition 40/60 at.%, examined *in situ* under pressure, the initial phase mixture Bi + Sn was transformed to the intermediate phase after annealing the sample at 2.9 GPa and 150° C for 1 hour. The diffraction pattern observed after annealing corresponded to the *oC*16 phase. The superconducting transition temperature ($T_c$) in BiSn alloy at ambient pressure is 3.7 K, a characteristic of β-Sn in the phase mixture with nonsuperconducting Bi. Under pressure, after the formation of the intermediate phase on the annealing at 363 K and 3.0 GPa, an increase in $T_c$ to 7.4 K was observed (see Ref. [18] and refs. therein).

On the whole, in all three Bi-based systems intermediate phases formed under pressure-temperature conditions from two-phase mixtures have the same *oC*16 structure identical to high pressure – high temperature form of Bi and these phases can be considered as extended solid solutions based on Bi- *oC*16. In addition an atomic ordering within 8f and 8d positions was observed for the nearly equiatomic BiSn-*oC*16 phase. Structural resemblance of Bi-based phases is supplemented by similar superconducting properties that have its source from the discussed Hume-Rothery mechanism.

## 3. Method of Analysis

The underlying concept of structural stability in the nearly-free electron model is based on matching of the Fermi sphere (FS) radius, $k_F$, and the distance to the Brillouin zone plane, $½q_{hkl}$, that is half of the reciprocal vector $q_{hkl}$, $k_F \approx ½ q_{hkl}$ [7,8]. This rule defines boundaries of phase stability by average number of valence electrons per atom for classical Hume-Rothery phases such as fcc, bcc and hcp. For complex phases such as γ-brass, $Cu_5Zn_8$ - *cI*52 it is necessary to consider the extended zone or Brillouin-Jones zone, assigned here as BZ. For complex phases it is important how well the BZ accommodates the FS which provides the gain in band structure energy to be enough to compensate the loss in electrostatic energy that arises from the formation of superlattices, vacancies and distortions.

To analyze complex phases within the FS – BZ interactions the program BRIZ is used that allows to consider high symmetry structures, cubic or hexagonal, as well as structures of low symmetry: tetragonal, orthorhombic, monoclinic or triclinic [22,36]. As an input the following data are used: lattice parameters, number of atoms in the cell, number of valence electrons per atom, and the hkl indices of the Brillouin planes that are selected from peaks in the diffraction pattern located near $k_F$. The output data are the BZ volume, the FS volume, $k_F$, and distances to the BZ planes. The program gives BZ construction with the inscribed FS as shown in Figure 2. This approach allows qualitative estimations of stability of structure and some physical properties of materials with this structure.

Thus, with the FS-BZ consideration for the phases in compressed alkali metals Li and Na, the appearance of a complex *cI*16 structure after bcc and fcc is evident to be of electronic origin of Hume-Rothery type, when superlattice 2x2x2 is formed due to a slight atomic shift to produce a Bragg plane close to $k_F$, at that Li and Na being monovalent [2,11]. When considering the next phase in Na with the *oP*8 structure it is essential to assume Na being divalent as for the isostructulal phase AuGa [11].

The same FS-BZ analysis can be applied to the problem of incommensurate structures by using suitable approximants as shown for incommensurately modulated phase of phosphorous at pressures 100-137 GPa [41]. Incommensurate host-guest phases are found in alkali metals Na, K, Rb and they can be considered in relation to the Hume-Rothery mechanism with the proper approximant extracting the necessary valence electron count from the FS – BZ model as well as Bi-III host-guest phase [42]. It is important to mention that these host-guest phases in alkali elements are in line among the other high-pressure phases that are suggested to be considered within Hume-Rothery mechanism, as the *oC*16 phase discussed in this paper.

## 4. Conclusions

In this paper the *oC*16-*Cmca* structure is analyzed for the purpose of understanding the formation and stability of this structure in a number of elements from the both sides of the Periodic Table as well as in some binary alloys as for instance in Bi alloys with In, Pb and Sn. Primarily, the *Cmca-oC*16 structure is a common high pressure form for polyvalent metals and alloys with the average number of valence electrons ~4, and might well be found in other alloys with the constituents from groups III-VI under pressure. Stability of this rather complex phase is related to the Hume-Rothery mechanism and gaining in the valence electron energy. Construction of Brillouin zone boundaries with the Bragg plains close to the Fermi sphere results in the highly-symmetrical polyhedron with many faces contacting the Fermi sphere and nearly filled by electron states. This configuration meets well the Hume-Rothery mechanism criteria for phase stability explaining qualitatively the physical properties of the matter with this structure such as resistivity and superconductivity.

Unexpected formation of the *oC*16-*Cmca* structure in compressed alkali elements K, Rb and Cs brings up the question of the valence electron level hybridization with the upper core electrons. The commonly accepted $s - d$ electron transition is sufficient to explain the first few high-pressure transitions in alkali metals and on further compression it is necessary to extend the degree of electron level hybridization that may account for sequences of complex high pressure phases including discussed in this paper the *oC*16 phase.

## Acknowledgments

The author gratefully acknowledges Dr. Olga Degtyareva for valuable discussion and comments. This work is supported by the Program of Russian Academy of Sciences "The Matter under High Energy Density".

## References

1. McMahon, M.I; Nelmes, R.J. High-pressure structures and phase transformations in elemental metals. *Chem. Soc. Rev.* **2006**, *35*, 943-963.
2. Degtyareva, V.F. Simple metals at high pressures: The Fermi sphere - Brillouin zone interaction model. *Physics-Uspekhi* **2006**, *49*, 369-388.
3. Degtyareva, O. Crystal structure of simple metals at high pressures. *High Press. Res.* **2010**, *30,* 343–371.
4. Schwarz, U.; Takemura, K.; Hanfland, M.; Syassen, K. Crystal structure of cesium-V. *Phys. Rev. Lett.* **1998**, *81*, 2711-2714.
5. Hanfland, M.; Schwarz, U.; Syassen, K.; Takemura, K. Crystal structure of the high-pressure phase silicon VI. *Phys. Rev. Lett.* **1999**, *82*, 1197-1200.
6. Schwarz, U.; Jepsen, O.; Syassen, K. Electronic structure and bonding in the Cmca phases of Si and Cs. *Solid State Commun.* **2000**, *113*, 643-648.

7. Mott, N.F.; Jones, H. *The Theory of the Properties of Metals and Alloys,* London: Oxford University Press, 1936.
8. Jones, H. *The Theory of Brillouin Zones and Electron States in Crystals*, Amsterdam: North-Holland, 1962.
9. Hume-Rothery, W. *Atomic Theory for Students of Metallurgy,* London: Institute of Metals, 1962.
10. Ross, M.; McMahan, A.K. Systematics of the $s{\rightarrow}d$ and $p{\rightarrow}d$ electronic transition at high pressure for the elements I through La. *Phys. Rev. B* **1982**, *26*, 4088-4093.
11. Degtyareva, V.F.; Degtyareva, O. Structure stability in the simple element sodium under pressure. *New J. Phys.* **2009**, *11*, 063037.
12. Pearson, W.B. *Crystal Chemistry and Physics of Metals and Alloys*, Wiley-Interscience, New York, 1972.
13. Kubiak, R.; Wolcyrz, M. Refinement of crystal structures of $AuSn_4$ and $PdSn_4$. *J. Less-Common Met.,* **1984**, *97*, 265-269.
14. Takemura, K.; Schwarz, U.; Syassen, K.; Hanfland, M.; Christensen, N.; Novikov, D.; Loa, I. High-pressure Cmca and hcp phases of germanium. *Phys. Rev. B* **2000**, *62*, R10603–R10606.
15. Schwarz, U.; Syassen, K.; Grzechnik, A; Hanfland, M. The crystal structure of rubidium-VI near 50 GPa. *Solid State Commun.* **1999**, *112*, 319–322.
16. Lundegaard, L.F.; Marqués, M.; Stinton, G.; Ackland, G.J.; Nelmes, R.J.; McMahon, M.I. Observation of the oP8 crystal structure in potassium at high pressure. *Phys. Rev. B* **2009**. *80*, 02010.
17. Chaimayo, W.; Lundegaard, L.F.; Loa, I.; Stinton, G.W.; Lennie, A.R.; McMahon, M.I. High-pressure, high-temperature single-crystal study of Bi-IV. *High Press. Res.* **2012**, *32*, 442-449.
18. Degtyareva, V.F.; Degtyareva, O; Allan, D.R.. Ordered Si-VI-type crystal structure in BiSn alloy under high pressure. *Phys. Rev. B* **2003**, *67*, 212105.
19. Degtyareva, V.F. Crystal structure of a high-pressure phase in Bi-based alloys related to Si-VI. *Phys. Rev. B* **2000**, *62*, 9-12.
20. Kittel, C. *Introduction to Solid State Physics,* JohnWiley & Sons, Canada, 1995.
21. Shannon, R.D. Revised Effective Ionic Radii and Systematic Studies of Interatomic Distances in Halides and Chaleogenides. *Acta Cryst. A* **1976**, *32*, 751-767.
22. Degtyareva, V.F.; Smirnova, I.S. BRIZ: a vizualization program for Brillouin zone – Fermi sphere configuration. *Z. Kristallogr.* **2007**, *222*, 718-721.
23. Sato, H.; Toth, R.S. Fermi Surface of Alloys. *Phys. Rev. Lett*. **1962**, *8*, 239–241.
24. Harrison, W.A. *Pseudopotentials in the Theory of Metals*, Elsevier, New York, 1966.
25. Bundy, F.P. Phase diagram of bismuth to 130,000 $kg/cm^2$, 500°C. *Phys. Rev.* **1958**, *110*, 314-318.
26. McMahon, M.I.; Degtyareva, O.; Nelmes, R.J. Ba-IV-type incommensurate crystal structure in group-V metals, *Phys. Rev. Lett.* **2000**, *85*, 4896–4899.
27. Ponyatovskii, E.G; Degtyareva, V.F. Specific Features of T-C-P diagrams for binary systems of B-elements. *High Press. Res.* **1989**, *1*,163-184.
28. Mizutani, U. *Introduction to the Electron Theory of Metals,* Cambridge University Press, 2001.

29. Mizutani, U. *Hume-Rothery rules for structurally complex alloy phases,* CRC Press, Taylor & Francis, London, 2011.
30. Häussler, P. Interrelations between atomic and electronic structures - Liquid and amorphous metals as model systems. *Physics Report* **1992**, *222*, 65-143.
31. Stiehler, M.; Rauchhaupt, J.; Giegengack, U.; Häussler, P. On modifications of the well-known Hume-Rothery rules: Amorphous alloys as model systems. *J. Non-Cryst. Solids*, **2007**, *353*, 1886-1891.
32. von Schnering, H.G.; Nesper, R. How nature adapts chemical structures to curved surfaces. *Angew. Chem. Int. Ed. Engl.* **1987**, *26*, 1059–1080.
32. Nesper, R; Grin, Yu. Periodic Space Partitioners (PSP) and their relations to crystal chemistry. *Z. Kristallogr.* **2011**, *226*, 692-710.
34. Ma, Y.; Eremets, M.;, Oganov, A.R.; Xie,Y.; Trojan, I.; Medvedev, S.; Lyakhov, A.O.; Valle, M.; Prakapenka, V. Transparent dense sodium. *Nature* **2009**, *458*, 182-186
35. Marqués, M.; Ackland, G.J.; Lundegaard, L.F.; Stinton, G.; Nelmes, R.J.; McMahon, M.I. Potassium under pressure: A pseudobinary ionic compound. *Phys. Rev. Lett.* **2009**, *103*, 115501.
36. Degtyareva, V.F.; Afonikova, N.S. Simple metal binary phases based on the body centered cubic structure: Electronic origin of distortions and superlattices. *J. Phys. Chem. Solids* **2013**, *74*, 18-24.
37. Wittig, J. Pressure-Induced Superconductivity in Cesium and Yttrium. *Phys. Rev. Lett.* **1970**, *24*, 812-815.
38. Erskine, D.; Yu, P.Y.; Chang, K.J.; Cohen, M.L. Superconductivity and Phase Transitions in Compressed Si to 45 GPa. *Phys. Rev. Lett.* **1986**, *57*, 2741-2744.
39. Hirsch, J.E.; Hamlin, J.J. Why non-superconducting metallic elements become superconducting under high pressure. *Physica C* (Suppl.1) **2010**, *470*, S937-939.
40. Hirsch, J.E. Materials and mechanisms of hole superconductivity. *Physica C* **2012**, *472*, 78-82.
41. Degtyareva, V.F. Electronic origin of the incommensurate modulation in the structure of phosphorus IV. *J. Phys.: Conf. Ser.* **2010**, *226*, 012019.
42. Degtyareva, V.F.; Degtyareva, O. Potassium under pressure: electronic origin of complex structures. (To be published).